\documentclass[12pt]{article}

\usepackage[utf8]{inputenc}
\usepackage{amsmath}
\usepackage{charter}
\usepackage{fullpage}
\usepackage{graphicx}
\usepackage[sort&compress,numbers]{natbib} 
\usepackage{hyperref}
\hypersetup{hidelinks}
\usepackage[total={6.5in,9in},top=1in,headsep=0.1in,headheight=1in]{geometry}

\usepackage{multirow}

\newcommand\snowmass{
\begin{center}
  \rule[-0.2in]{\hsize}{0.01in}\\
  \rule{\hsize}{0.01in}\\
  \vskip 0.1in
  Submitted to the Proceedings of the US Community Study\\ 
  on the Future of Particle Physics (Snowmass 2021)\\
  \rule{\hsize}{0.01in}\\
  \rule[+0.2in]{\hsize}{0.01in}\\[-2em]
\end{center}
}

\usepackage[firstpage=true]{background}
\backgroundsetup{contents={\parbox{6.5in}{\snowmass}}, scale=1,placement=top,opacity=1,color=black,position={3.25in,1.2in}}

\usepackage{fancyhdr}
\fancypagestyle{plain}{%
  \fancyhf{}%
  \fancyhead[C]{}
  \fancyfoot[C]{\thepage}
}

\fancypagestyle{empty}{%
  \fancyhf{}%
  \fancyhead[C]{{\it Light Dark Matter Detection with Hydrogen-rich Crystals and Low-Tc TES Detectors}}
  \fancyfoot[C]{\thepage}
}
\title{Light Dark Matter Detection with Hydrogen-rich Crystals and Low-Tc TES Detectors}
\date{}

\usepackage{authblk}

\author[1]{Gensheng Wang}
\author[1,2,3]{Clarence L. Chang}
\author[1]{Marharyta Lisovenko} 
\author[4]{Valentine Novosad}
\author[1]{Volodymyr G. Yefremenko}
\author[1]{Jianjie Zhang}
 
\affil[1]{High Energy Physics Division, Argonne National Laboratory, Lemont, IL 60439, USA}
\affil[2]{Kavli Institute for Cosmological Physics, University of Chicago, Chicago, IL 60637, USA}
\affil[3]{Department of Astronomy and Astrophysics, University of Chicago, Chicago, IL 60637, USA}
\affil[4]{Materials Science Division, Argonne National Laboratory, Lemont, IL 60439, USA}

\begin{document}

\maketitle

\begin{abstract}
Direct detection of nuclear scatterings of sub-GeV Dark Matter (DM) particles favors low-Z nuclei. 
Hydrogen nucleus, which has a single proton, provides the best kinematic match to a light dark matter particle. 
The characteristic nuclear recoil energy is boosted by a factor of a few tens from those for larger nuclei used in traditional Weakly Interacting Massive Particle (WIMP) searches. 
Furthermore, hydrogen is optimal not only for spin-independent nuclear scatterings of sub-GeV DM, but also for spin-dependent nuclear scatterings, where large parameter space remains unconstrained.  
In this paper, we first introduce hydrogen-rich crystals, which include water ice, acetylene, anthracene, trans-stilbene, and naphthalene. 
These crystals emit two classes of signals under kinetic excitations. 
One class of the signals is infrared photons, which are from optically active fundamental vibrational modes of molecules and are at corresponding characteristic wavelengths. 
The other is acoustic phonons, and optical phonons that decay into acoustic phonons. 
We then discuss the technical status and future researches of low-Tc Transition-Edge Sensor (TES) detectors, which measure single infrared photons and a small flux of acoustic phonons with desirable sensitivities. 
With theoretical modeling to select the hydrogen-rich crystals for the optimized science reach, development of ultra-sensitive low-Tc TES detectors for readout, and experimentally characterizing transport properties of photons and phonons in the selected hydrogen-rich crystals, a direct detection experiment can be built for measuring the large unexplored parameter space of light DM particles.
\end{abstract}

\section{Introduction}
To identify the new physics of Dark Matter (DM), the Basic Research Needs for Dark Matter Small Project New Initiatives ~\cite{BRN:18} recommended to ``detect individual galactic dark matter particles below the proton mass through interactions with advanced, ultra-sensitive detector.'' 
The recommendation is driven by new theories, which include hidden-sector DM~\cite{Hodges:93, Feng:08, Pospelov:08, Boem:04, Foot:15, Knapen:17, Fabbrichesi:21}, asymmetric DM~\cite{Kaplan:09, Lin:12, Zurek:14}, freeze-in DM~\cite{Hall:10} and strong interacting DM~\cite{Hochberg:14, Emken:19}. These theories provide well-motivated DM candidates, which can have masses lighter than that of the traditional Weakly Interacting Massive Particles (WIMPs)~\cite{Jungman:96} and are beyond the science reaches of most second-generation experiments~\cite{Battaglieri:17, Schumann:19}.

The existing and proposed light DM direct detection projects~\cite{Battaglieri:17, Schumann:19} cover a large number of target materials with a variety of signal readout methods.  
Each of them has advantages and disadvantages.  
For example, superfluid $^4$He~\cite{Maris:17, Guo:13, Hertel:18, Schutz:16} is sensitive to spin-independent light DM nuclear scatterings. 
However, it is insensitive to spin-dependent scatterings because $^4$He has two protons and two neutrons and its net nuclear spin is zero.  
Crystalline solids are utilized with phonons~\cite{Griffin:18, Griffin:08, Kurinsky:19, Griffin:2020}, scintillation~\cite{Griffin:18, Griffin:2020}, ionization~\cite{Essig:16}, ionization with a thermal gain~\cite{Alkhatib:21}, or low threshold thermal measurement~\cite{Abdelhameed:19}.
However, the target nuclei interacting with DM are much heavier than the proton mass and therefore are not optimal for kinetic energy deposition from light DM due to kinematic mismatch.  
Therefore, most of the experiments and proposals using crystalline solids have been optimized to sub-GeV DM electron scatterings, not nuclear scatterings.  
Recently, there are proposals using hydrogen. 
However, traditional ionization and scintillation readout signals do not have low enough energy thresholds to explore the parameter space below a DM mass around 100 MeV. 
For example, the NEWS-G collaboration is searching for light DM candidates using a spherical proportional counter with hydrogen-rich gases, providing access in the 0.1–10 GeV mass range~\cite{Nikolopoulos:20}.  
For hydrogen-rich plastic scintillators~\cite{Collar:18, Blanco:20}, a large excitation energy is needed to create a visible photon on average. 
The scintillation light is from $\pi$-$\pi^*$ transitions due to alternating single and double bonds between their carbon atoms in molecules, not directly from C-H bonds with a low excitation energy. 

A hydrogen-rich crystal, which can be crystalline water ice with O-H bonds or selected hydrocarbon crystals with C-H bonds, emits mid-infrared photons and acoustic phonons when dark matter collides with a hydrogen nucleus.  
By operating hydrogen-rich crystals and low-Tc TES detectors at milli-Kelvin temperatures, these infrared photons and acoustic phonons can be measured with low-Tc TES detectors at low thresholds.
The approach that we propose to search for light DM particles has the following features:
\begin{enumerate}
    \item Uses hydrogen atoms with the lightest target nuclei. When a DM particle elastically scatters off a nucleus, the maximum kinetic energy deposition is $2\mu^2 v_{\chi}^2/m_N$, where $\mu=m_{\chi}m_N/(m_{\chi}+m_N)$ is the reduced mass, $m_{\chi}$ is the mass of the DM particle, $m_N$ is the target nucleus mass. $v_{\chi}$ is the escape velocity of the Milky Way galaxy plus the Earth velocity. Hydrogen nucleus, which has a single proton, provides the best kinematic matching in mass to a sub-GeV DM particle. 
    Therefore, the characteristic nuclear recoil energy is boosted by a factor of a few tens from those for larger nuclei used in the traditional WIMP searches.
    \item Takes advantage of the low excitation energy of molecules~\cite{Essig:19} and crystalline lattices~\cite{Griffin:2020}. For the hydrogen-rich crystals (such as water ice, acetylene, anthracene, trans-stilbene, and naphthalene) that will be discussed in this paper, their typical molecular vibrational excitation energy is on the order of 100 meV. Their optical phonon excitation energy is around a few tens of meV. Their acoustic phonon excitation energy is around a few meV. These are sharp contrasts to traditional ionization or scintillation of semiconductors and noble gases, which cost a few to a few tens eV of energy on average for a readout signal. 
    \item Uses ultra-sensitive low-Tc TES detectors for low threshold detection. Low-Tc TES detectors can have an energy resolution on the order of 10 meV with existing technologies and have potentials for further improvement with additional R\&Ds.
\end{enumerate}

In this paper, we describe hydrogen-rich crystals and their excitation signals when interacting with a DM particle in section~\ref{crystals}, review the technical status of low-Tc TES detectors with an update of our work and future researches in section~\ref{TES-Detectors}, and conclude in section~\ref{Conclusion}.  

\section{Hydrogen-rich Crystals for Light Dark Matter Detection}
\label{crystals}

\subsection{Crystalline Water Ice}
\label{waterice}
\begin{figure}[!ht]
	\centering
	\includegraphics[height=2.2in, keepaspectratio]{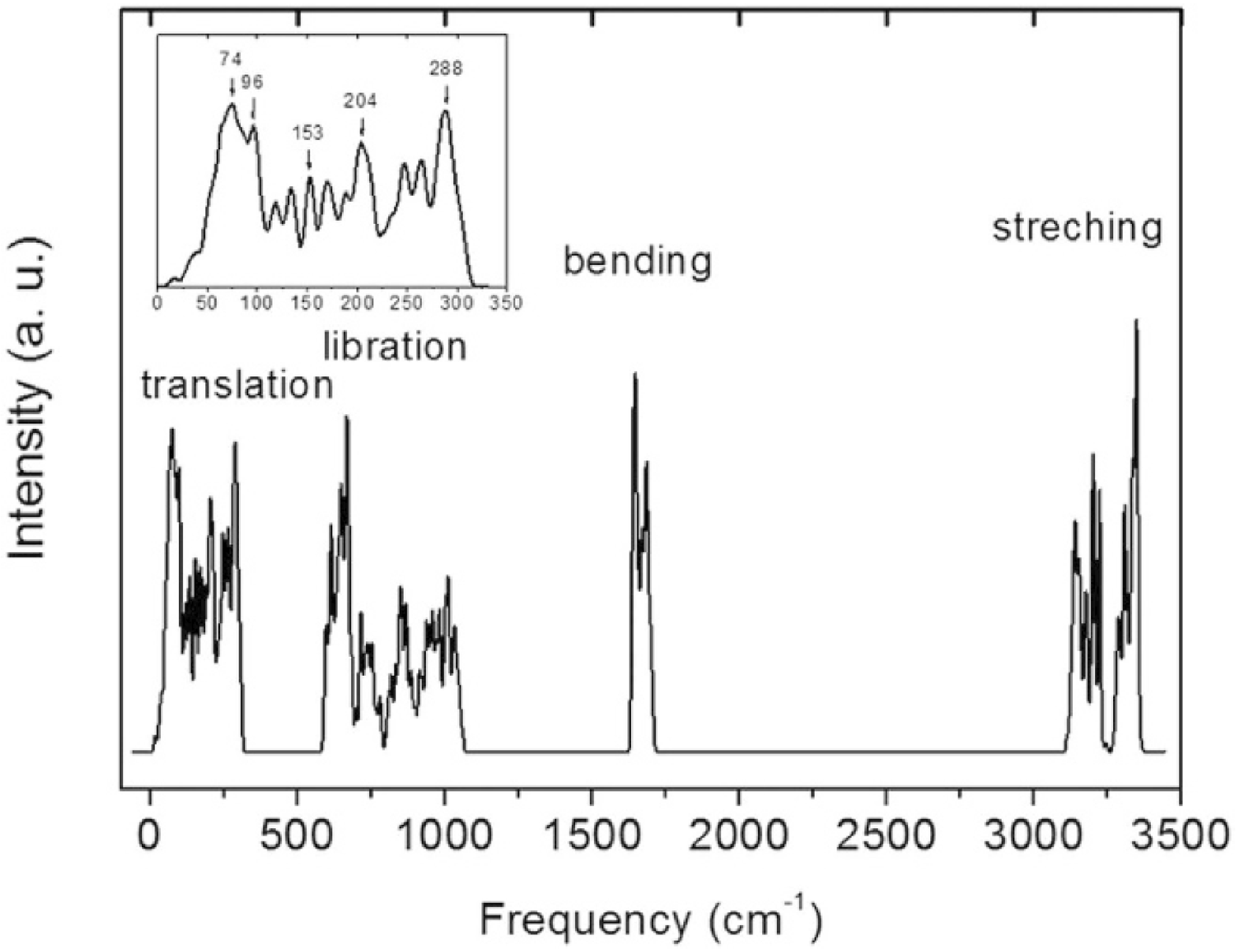}
	\includegraphics[height=2.2in, keepaspectratio]{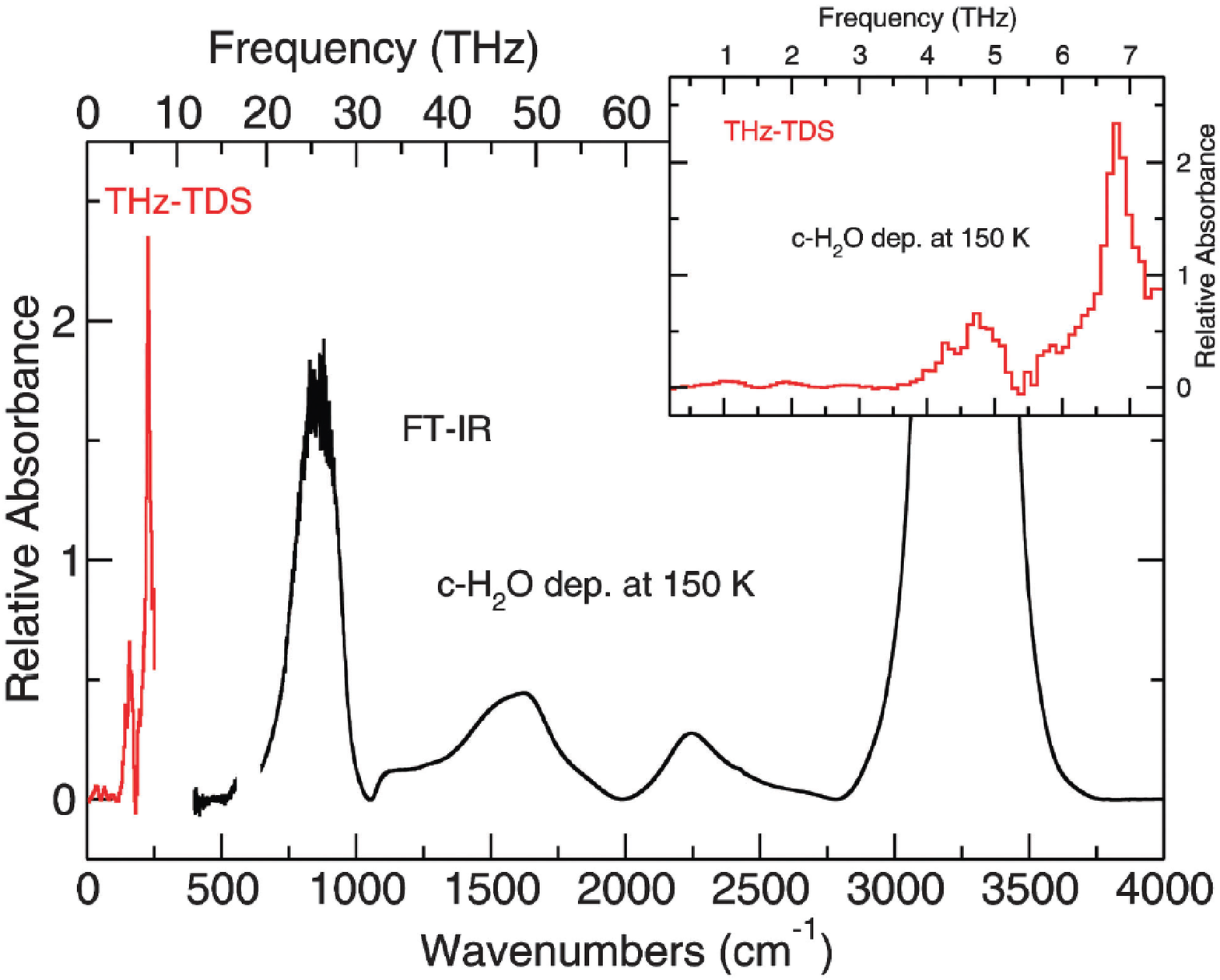}
	\caption{{\it Left}: The normal modes of crystalline water ice XI from Zhang {\it et al.}~\cite{Zhang:16}. {\it Right}: The measured infrared spectrum of crystalline water ice from Allodi {\it et al.}~\cite{Allodi:14}. The libration and translation modes are all optically active due to a net dipole moment of H$_2$O beyond the O-H bond dipole moments.}
	\label{fig:ice}
\end{figure}

Crystalline water ice has a rich infrared spectrum~\cite{Zhang:16, Wozniak:12, Liu:13, Allodi:14}.  
These infrared photons are from mid-infrared up to far-infrared in wavelength. 
They are mapped to the fundamental vibrational modes of H$_2$O molecules.
Figure~\ref{fig:ice} shows the calculated normal modes and measured infrared spectrum of crystalline water ice. 
The typical infrared emissions are: (1) intramolecular O-H vibrations emitting infrared photons at 2.94 $\mu$m (O-H asymmetric stretching), 3.11 $\mu$m (O-H symmetric stretching), and 6.17 $\mu$m (H-O-H bending);  
(2) the liberation mode of H$_2$O molecules emitting mid-infrared photons in a broad band centered at 11.7 $\mu$m; 
(3) the stretching and bending modes of intermolecular hydrogen bonds (O–H$\cdot\cdot\cdot$O) emitting infrared photons at 44 $\mu$m and 166 $\mu$m respectively. 
These infrared photons have been used to identity water ice in astronomy~\cite{Allodi:14, Dijkstra:06, Dudley:08, Loon:10, Malfait:99, Cohen:99}.

\begin{figure}[!hb]
    \centering
    \includegraphics[height=3in, keepaspectratio]{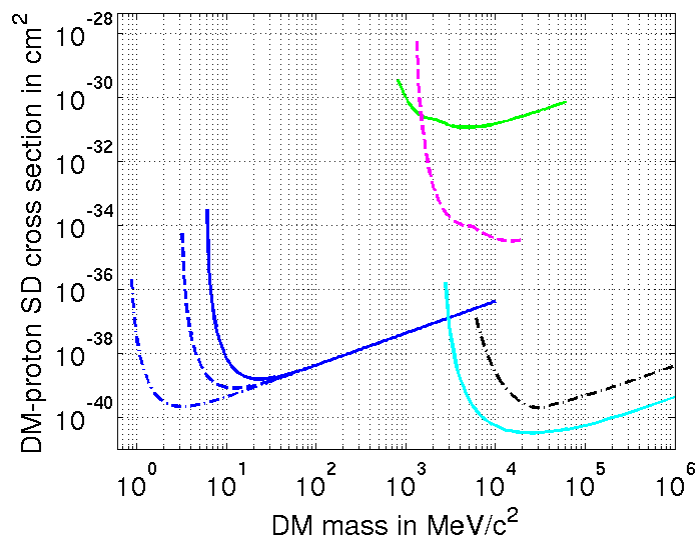}
    \caption{Exclusion limits (at 90\% CL) with 1000 g$\cdot$day crystalline water ice exposure for spin-dependent DM-proton scattering cross section. 
    Zero background is assumed. 
    Blue  solid: Ice with E$_{th}$=0.50 eV. 
    Blue dashed line: Ice with E$_{th}$=0.14 eV. 
    Blue dot dashed: Ice with E$_{th}$=0.01 eV. 
    Green solid: CRESST Li-7~\cite{Abdelhameed:19epj}. Magenta dashed: CDMSlite SD proton~\cite{Agnese:18prd}. 
    Black dot dashed: XENON1T 2019~\cite{Aprile:19prl}.
    Cyan solid: PICO-60 2019~\cite{Amole:19prl}.}
    \label{fig:limit2}
\end{figure}

The scattering of a DM particle off a hydrogen nucleus in water ice can be understood in two steps: 
(1) a DM particle interacts with a proton and transfers its energy and momentum to the proton; 
(2) the transferred energy and momentum excite vibrational modes of O-H bonds and liberation modes of H$_2$O molecules.  
Any excess energy and momentum are transferred to the center-of-mass frame of the molecule and to the neighbor molecules. 
Therefore, highly efficient DM kinetic energy to infrared photon conversion is expected when the deposited kinetic energy equals the energy of an optically active fundamental mode.
A detection of a single 11.7 $\mu$m infrared photon from the liberation mode can be sensitive to a DM particle with mass about 3 MeV. 
Furthermore, a water molecule has a large electric dipole moment (1.85 D). 
So, it is sensitive to photons.  
This makes water ice an ideal detection target for dark photon mediated electron scattering or nuclear scattering based on photon mixing theory~\cite{Knapen:17, Fabbrichesi:21, Alexander:16}. 
Moreover, a heavy dark photon itself can be a DM candidate, which interacts with a water molecule through the electric dipole moment.  
The 11.7 $\mu$m liberation mode will allow detection of a dark photon with mass about 33 keV.

The other primary excitations in crystalline water ice are acoustic phonons, and optical phonons that decay into acoustic phonons because of anharmonic potential between molecules~\cite{Levinson:80, Tamura:85, Maris:93}.
Furthermore, the primary infrared photons from de-excitation of optically active modes in a bulk ice crystal can be re-absorbed and turned into phonons through vibrational relaxation~\cite{Hill:88, Poulsen:03, Werhahn:12}.
Phonons in crystalline water ice have been experimentally investigated~\cite{Abe:11ice, Wehinger:14, Glebov:00, Strassele:04, Bennington:99} at temperatures of a few tens of Kelvin and theoretically studied~\cite{Adeagbo:05} with first principles based on density functional theory before. 
In crystalline water ice, the low energy acoustic phonons have a linear dispersion relation, propagate at sound speed, and are in sub-terahertz. 
These acoustic phonons scatter less than photons do because of sparse excitation states below 1THz~\cite{Zhang:16, Allodi:14} at an operation temperature of 10 mK. 
Therefore, a large acoustic phonon mean free path is expected. 
These acoustic phonons can be measured with TES detectors as what has been done with silicon target in the CDMS experiments~\cite{Cabrera:00, Saab:02}.
The advantage to measure phonons is that large size (1 cm or larger) water ice crystals can help scale up the detection target mass.

The spin-dependent DM-nucleon scattering cross sections for crystalline water ice can be estimated by using the method~\cite{Jungman:96, Abdelhameed:19epj, Agnese:18prd, Bringmann:17, Lewin:96} developed in the DM direct search community.
The blue lines in Figure~\ref{fig:limit2} are the potential science reaches at three different energy thresholds by measuring photons or phonons with TES detectors at milli-Kelvin temperatures. 
The use of hydrogen target for nuclear scattering and low threshold TES detectors for readout allows to study the unexplored parameter space of light DM particles with a few grams of water ice in one year detection time.  

Note that there was a proposal to use crystalline water ice for DM detection~\cite{Vavra:14} before. Limited by detector technology, the proposed readout signal was optical photons, which are the overtones of the O-H stretching. 
The consequence is that the measured event rate can be suppressed by several orders of magnitude. 
In the current proposal, infrared photons and acoustic phonons, which can be read out with TES detectors at high efficiency, are from the fundamental modes activated directly with deposited kinetic energy and momentum. 
Therefore, there is no suppression of event rate. 

Note that high quality single crystal water ice has been grown successfully before with a variety of methods. Knight~\cite{Knight:96} developed a method similar to the natural, lake grown ice process.  
Khusnatdinov and Petrenko~\cite{Khusnatdinov:96} developed a novel technique utilizing a vacuum chamber that can produce large “plate” like ice crystals. 
The zone refining technique as described by Bilgram {\it et al.}~\cite{Bilgram:73} has been used to make high quality crystalline ice. 
More recently, Bisson {\it et al.}~\cite{Bisson:16} reliably produced relatively large water single crystals that are optically flawless with the Stockbarger modified Bridgeman technique. 

\subsection{Crystalline hydrocarbons}
\label{hydrocarbons}
Crystalline hydrocarbons are another class of candidate materials suitable using hydrogen nuclei for light DM detection.  
The signals for measurement after a DM particle colliding with a hydrogen nucleus are infrared photons and acoustic phonons.
Although there is no net electric dipole in a hydrocarbon molecule, the C-H bonds have large bond dipole moments, which result in large change rates of the dipole moments when fundamental vibrational modes of the molecule are excited~\cite{Wingfield:55, Eggers:55, Cole:64}.
Therefore, they are infrared emission materials under kinetic energy excitation.  
The emitted infrared photon frequencies depend on the molecular geometry, atomic masses, and intramolecular forces in the material.  
The emission (or absorption) intensities depend on the square of the change rates of dipole moments from the molecular vibrational displacements specified by normal coordinates.
As is in inorganic molecular solids, phonons in crystalline hydricarbons are the quanta of the molecule lattice vibrations~\cite{Schwoerer:07}.

There are a large number of hydrocarbons. 
We only describe a few examples below for their simplicity or maturity of materials synthesis. 
Acetylene, which has a H-C$\equiv$C-H linear symmetric molecular structure and a well-understood infrared spectrum, can be a potential candidate target for light DM detection. 
Other candidates include simple polycyclic aromatic hydrocarbons, such as crystalline anthracene, trans-stilbene, and naphthalene. 
These three have simple molecular structures with only three or two aromatic rings, and have been widely studied for scintillation applications.

\subsubsection{Acetylene}
\label{acetylene}
Acetylene, C$_2$H$_2$, is a well-understood simple molecule with C-H bonds and high Debye temperature. Crystalline acetylene has two known phases. One is the high temperature cubic phase (CP), which is stable between 133 K and the melting point of 191 K. The other is the low temperature orthorhombic phase (OP), which is stable below 133 K. Typically, the OP phase is formed by condensing acetylene gas at a slow rate at the liquid nitrogen temperature. 

\begin{table}[!ht]
\begin{center}
\caption{Calculated and observed frequencies in cm$^{-1}$ of optically active vibration modes of crystalline acetylene in orthorhombic phase. For the five internal modes, $\nu_1$ is due to symmetric C-H stretching, $\nu_2$ is due to C-C stretching, $\nu_3$ is due to asymmetric C-H stretching, $\nu_4$ is due to molecular trans-bending, and $\nu_5$ is due to molecular cis-bending. Data are from reference~\cite{Binbrek:92}.} 
\label{tab:acetylence}
\begin{tabular}{ |c|c|c|c| } 
\hline \hline
Vibration Type & Symmetry Species & Calculated Frequency & Observed Frequency \\
\hline 
\multirow{2}{*}{Internal mode $\nu_1$} & $A_g$ & 3316.5 & 3324 \\ 
& $B_{3g}$ & 3316.1 & 3315 \\ 
\hline
\multirow{2}{*}{Internal mode $\nu_2$} & $B_{3g}$ & 1977.9 & 1961 \\ 
& $A_g$ & 1976.3 & 1951 \\ 
\hline
\multirow{2}{*}{Internal mode $\nu_3$} & $B_{2u}$ & 3233.5 & 3226.5 \\ 
& $B_{1u}$ & 3233.1 & 3226.5 \\ 
\hline
\multirow{4}{*}{Internal mode $\nu_4$} & $B_{2g}$ & 656.7 & 659.3 \\ 
& $A_g$ & 637.7 & 638.4 \\ 
& $B_{1g}$ & 634.3 & 628.7 \\ 
& $B_{3g}$ & 633.9 & - \\ 
\hline
\multirow{4}{*}{Internal mode $\nu_5$} & $A_u$ & 783.6 & - \\ 
& $B_{2u}$ & 756.8 & 768.8 \\ 
& $B_{1u}$ & 755.1 & 760.1 \\ 
& $B_{3u}$ & 755.0 & 747.5 \\ 
\hline
\multirow{3}{*}{Translation} & $B_{2u}$ & 127.5 & 127 \\ 
& $B_{1u}$ & 100.6 & 106 \\ 
& $A_u$ & 86.3 & - \\ 
\hline
\end{tabular}
\end{center}
\end{table}

For light DM detection, one signal channel is infrared photons. The deposited kinetic energy and momentum from a DM particle generates vibrations within the the acetylene molecules and between the molecules. 
The intramolecule and translation vibrational modes are optically active because of the large changes of electric dipole moments~\cite{Wingfield:55, Eggers:55}.  
The emission (or absorption) intensity of each mode is proportional to the square of the change rate of the electric dipole moment when the molecule is deformed in the manner specified by a normal coordinate. 
Table~\ref{tab:acetylence} summarizes the optically active modes of the crystalline acetylene in the OP phase. 
The intensities of the near- and mid-IR spectra of crystalline acetylene, which are from the fundamental internal molecule modes and their overtones, are shown in the Fig. 1 in ref.~\cite{Hudson:14}.
The cis-bending $\nu_5$ at 756 cm$^{-1}$ and the asymmetric CH stretching $\nu_3$ at 3233 cm$^{-1}$ are two major IR photon emission modes. 
The overtones of the internal vibrational modes at higher frequencies are much weaker.    

Another signal channel is phonons. 
Based on the orthorhombic crystalline structure, the classical atom-atom potential model between molecules, and the coulombic interactions generated by electric monopoles and multipoles situated on atom and bond sites~\cite{Marchi:85,Gamba:82,Leech:93}, the phonons in orthorhombic crystalline acetylene have been investigated theoretically.
The models are verified with the measured intensities of infrared spectra and sublimation energy of the acetylene crystal.
The calculated dispersion curves of orthorhombic acetylene crystal include well-defined optical and acoustic phonon branches. 
The acoustic phonons with linear dispersion relations~\cite{Marchi:85,Gamba:82} at small wavevectors are in sub-terahertz range. 
These acoustic phonons are expected to have a large mean free path. 
Therefore,  they are suitable for measurement using TES detectors or TES detectors with additional large area phonon collection fins~\cite{Cabrera:00,Saab:02}.

\subsubsection{Polycyclic Aromatic Hydrocarbons}
\label{aromatic}
Polycyclic aromatic hydrocarbons (PAHs) have more than one aromatic rings and are rich with C-H bonds. 
PAHs are known to have a mid-infrared emission spectrum~\cite{Sandfors:04, Draine:07,Bouwman:11, Boersma:09, Brenner:92, Tienlens:13}. 
They have been used in astronomical observations with a well-known family of emission lines at 3.3 $\mu$m (C-H stretching), 5.25 and 5.75 $\mu$m (combination of C-H bending and C-C stretching), 6.2 and 7.6 $\mu$m (C-C stretching), 8.6 $\mu$m (C-H in plane bending), and 11.2 – 14.1 $\mu$m (C-H out of plane bending). 
There are more detailed information regarding the infrared spectra of PAHs in literature. 
The table I in ref.~\cite{Boersma:09} summarizes the peak positions and associated modes of the infrared emission features. 
The Fig. 7 in ref.~\cite{Tienlens:13} highlights the mid-infrared spectra associated with vibrational modes of PAH molecules.
For light DM detection, the mid-infrared photons can be emitted with the deposition of kinetic energy and momentum when a DM particle collides with a hydrogen nucleus in a PAH molecule. 
 
Among the large number of polycyclic aromatic hydrocarbons, anthracene (C$_{14}$H$_{10}$), trans-stilbene (C$_{14}$H$_{12}$), and naphthalene (C$_{10}$H$_{8}$) have only three or two aromatic rings. 
They have been studied and utilized as organic scintillators~\cite{Aaviksoo:82, Priestley:68, Chaudhuri:69, Erememko:88, Fraboni:16, Balamurugan:07} in the optical wavelengths.
Therefore, they are well understood organic crystals with developed fabrication methods and laboratory measurement data.
Anthracene crystals can be prepared with solution growth~\cite{Zhang:099, Li:079} or Bridgeman technique~\cite{Arulchakkaravarthi:02}. 
Anthracene is a planar-conjugated hydrocarbon with a P2$_1$/a space group crystal structure. 
In each base centered monoclinic unit cell, there are two anthracene molecules~\cite{Cruickshank:56}.
Not only its infrared spectra~\cite{Bree:68, Mackie:15} but also its phonon dispersion curves~\cite{Broude:78, Dorner:78} were measured in lab and are well understood. 
Trans-stilbene crystals can be prepared with solution growth~\cite{Fraboni:16, Zaitseva:15} or Bridgeman technique~\cite{Arulchakkaravarthi:01}. 
It is a hydrocarbon hosting two phenyl groups joined by an ethylene bridge in the trans configuration (trans-1,2-diphenylethylene). 
The molecule (C$_6$H$_5$-CH=CH-C$_6$H$_5$) is approximately planar in the monoclinic crystalline state and has a C$_{2h}$ molecular symmetry~\cite{Hoekstra:75}.
Its infrared spectrum~\cite{Meic:78, Pecile:69, Watanabe:02, Govindan:02} were measured and are well understood. 
There is no report regarding the measured phonon dispersion curves of crystalline trans-stilbene yet.
However, its optical and acoustic phonons~\cite{Saito:96} were analyzed with the crystalline structure and atom–atom potential approximation. 
There are low frequency intramolecular torsional modes~\cite{Bree:80} of the phenyl groups with energy close to that of lattice vibrational modes.
These intramolecular twisting angles are small in solid state~\cite{Hoekstra:75} at low temperatures.
From the comparison of the calculated dispersion relations for the flexible (rotations with two degrees of freedom) and rigid molecular models~\cite{Saito:96}, the intramolecular twisting modes are divided into two groups. 
One group has a frequency of around 50 cm$^{-1}$ with little dispersion. 
The other group lies above 100 cm$^{-1}$ and shows a rather large dispersion. 
Both of them are optical phonons, which decay into acoustic phonons.
The acoustic phonons, which can propagate in the crystal at small wavevectors, are not affected with the intramolecular twisting degrees of freedom.  
Naphthalene crystals are typically prepared with Bridgeman techniue~\cite{Selvakumar:05, Suthan:10}.
Crystalline naphthalene has a structure with the P2$_1$/a space group, and with two molecules in the monoclinic unit cell, each situated at inversion centers\cite{ Brown-Altvater:16}.
Both the infrared spectrum~\cite{Mackie:15, Pimentel:52, Chakraborty:16} and the phonon dispersion curves~\cite{Broude:80, Natkaniec:80} of naphthalene were measured in laboratory and are well understood~\cite{Fedorov:15}.

Another signal channel for light DM detection with aromatic hydrocarbon crystals is phonons. 
Similar to the kinetic processes in crystalline water ice, the phonons come from two sources.
One is primary acoustic and optical phonons~\cite{Levinson:80, Broude:78} generated directly from an external excitation.
Another is the secondary phonons from the relaxation of molecular vibrations~\cite{Hill:88, Poulsen:03, Werhahn:12} in optical active modes.
The whole physical process can be understood as below. 
The deposited kinetic energy in crystalline aromatic hydrocarbons leads to production of molecular vibrations~\cite{Bree:68, Mackie:15, Meic:78, Pecile:69, Pimentel:52}. 
The lifetime of a molecular vibration in a crystal is short (\textless 1 ns)~\cite{Levinson:80}.
In this period of time,  either an infrared photon is emitted through the electronic state de-excitation, or the vibrational energy is transferred from the electronic system into the phonon one. 
Moreover, in case of a bulk hydrocarbon crystal, there is a chance that an infrared photon is re-absorbed and converted to phonons through molecular vibration relaxation.
In such a relaxation process, high-frequency optical phonons are produced first. 
Then the high-frequency optical phonons are converted into low-frequency acoustic phonons through a process called anharmonic decay~\cite{Levinson:80, Tamura:85, Maris:93}. 
The acoustic phonons have frequencies on the order of 1 THz or less. 
All the processes of relaxation up to the acoustic phonon appearance are localized near the spot of the kinetic energy deposition, since the group velocities of optical phonons are small and the relaxation and decay processes proceed very fast.
Acoustic phonons travel at sound speed~\cite{Levinson:80, Maris:90, Shields:93}.
The acoustic phonons can be measured with TES detectors or TES detectors with large area superconducting phonon collection fins~\cite{Cabrera:00, Saab:02}. 

\section{Low-Tc TES Detectors}
\label{TES-Detectors}
Measurements of single infrared photons in wavelengths from a few to a few tens $\mu$m and a small flux of athermal phonons (which are long-lived acoustic phonons) are required for using hydrogen-rich targets to detect light DM.  
Transition-Edge Sensor (TES) detectors, which can be read out in large number of pixels with multiplexing technologies~\cite{Bender:20, Doriesel:16, Dober:21},  are a developed technology to realize such ultra-sensitive measurements. 
TES detectors have been utilized to detect single photons for communication~\cite{Adriana:08, Schmidt:18, Fortsch:15} and for axion-like particle detection~\cite{Bastidon:15}.  
A TES detector measuring athermal phonons~\cite{Cabrera:00, Saab:02} was originally developed in the CDMS collaboration. It continues to be the science enabling technology~\cite{Alkhatib:21} in the SuperCDMS collaboration. 
Therefore, the well-understood TES detectors are a natural choice to measure single infrared photons and a small flux of athermal phonons.

For a TES detector operated under negative electro-thermal feedback, its expected energy resolution~\cite{Irwin:95} is 
\begin{equation} 
\Delta E \approx 2.35 \sqrt{\frac{4k_B T_c^2 C}{\alpha} \sqrt{\frac{n}{2}}}, 
\label{eq:resolution} 
\end{equation} 
where $k_B$ is the Boltzmann constant, $T_c$ is the TES transition temperature, $C$ is the TES heat capacity, and $n$ is an index around five determined by electron-phonon decoupling~\cite{Giazotto:06}. 
Here $\alpha \approx (T/R)(dR/dT)$ is a parameter characterizing the TES transition profile, where $R$ is the temperature-dependent resistance of the TES in the transition. 
For a TES made of metal films, its heat capacity is linearly dependent on $T$ at low temperatures. 
Therefore, the energy resolution in equation~\ref{eq:resolution} scales with $T^{3/2}$. 
More generally, equation~\ref{eq:resolution} states that a TES detector favors a low-$T_c$, a large $\alpha$, and a small heat mass (through $C$). 
Following the statement, we review the technical paths of TES detectors measuring single infrared photons and athermal phonons.

Lowering the $T_c$ of TES is an effective way to improve energy resolution. 
In the CRESST collaboration, a W TES with 15 mK $T_c$ was used with a 23.6 g CaWO$_4$ detector to realize an energy threshold as low as 38.1 eV~\cite{Abdelhameed:19}. 
Recently, Fink {\it et al.}~\cite{Fink:20} demonstrated that a 400 $\mu m$ $\times$ 100 $\mu m$ $\times$ 40 $nm$ W TES with a $T_c$ of 40 mK has a resolution of 40 meV, which makes 100 meV detection threshold possible.
The resolution can be further improved with a lower $T_c$ TES.   
So, reproducible fabrication of W films with $T_c$ below 40 mK and with sharp transition profile is an active area of research using sputtered W films~\cite{Abdelhameed:20}.
On the other hand, a low-$T_c$ TES utilizing proximity effect is an alternative technical path forward. 
It was demonstrated that Ir-based proximity films, which include both Ir/Pt bilayers and Au/Ir/Au trilayers~\cite{Hennings:20} sputtered at room temperature, have tunable $T_c$ down to 20 mK.

\begin{figure}[!htb]
	\centering
	\includegraphics[height=2.0in, keepaspectratio]{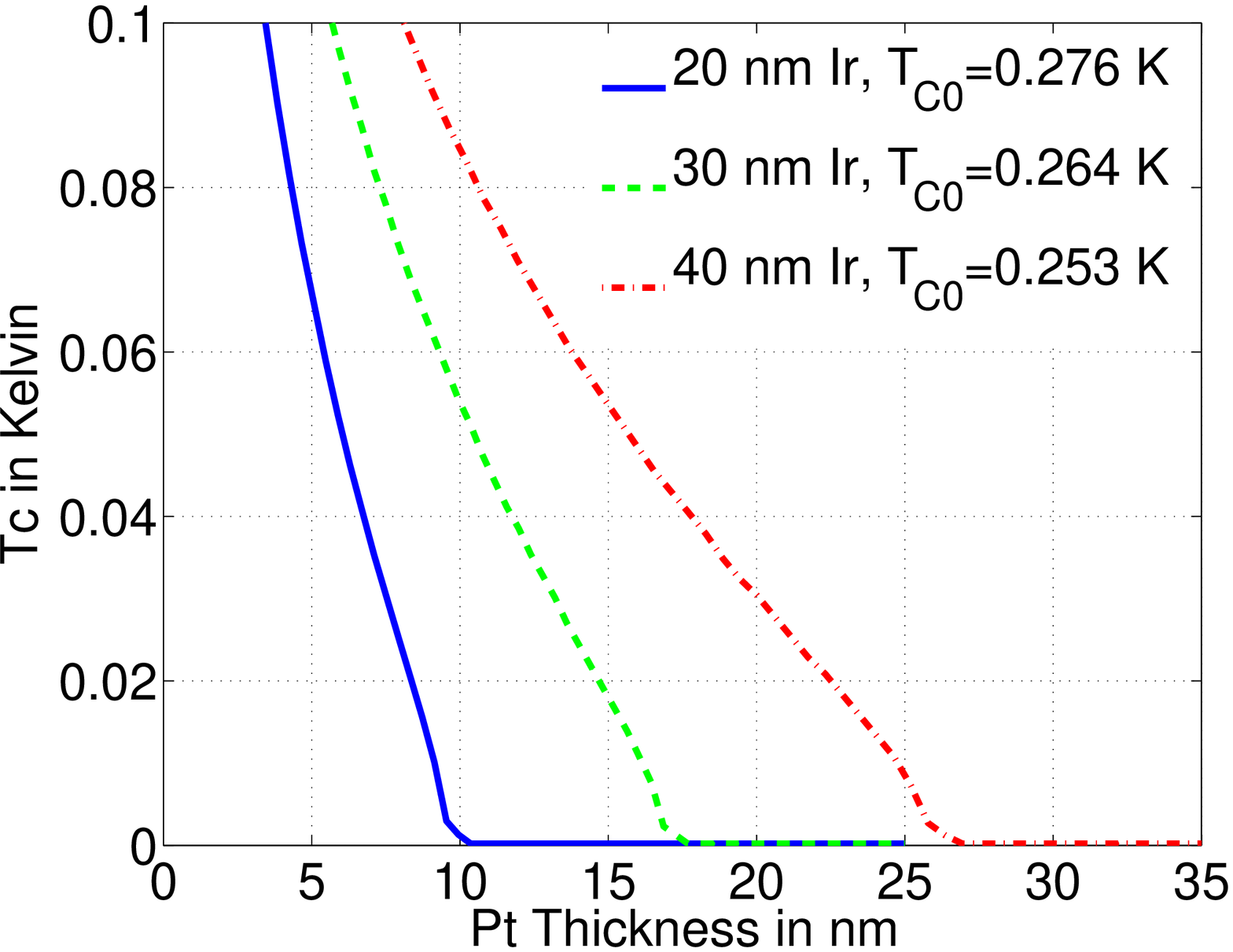}
	\includegraphics[height=2.0in, keepaspectratio]{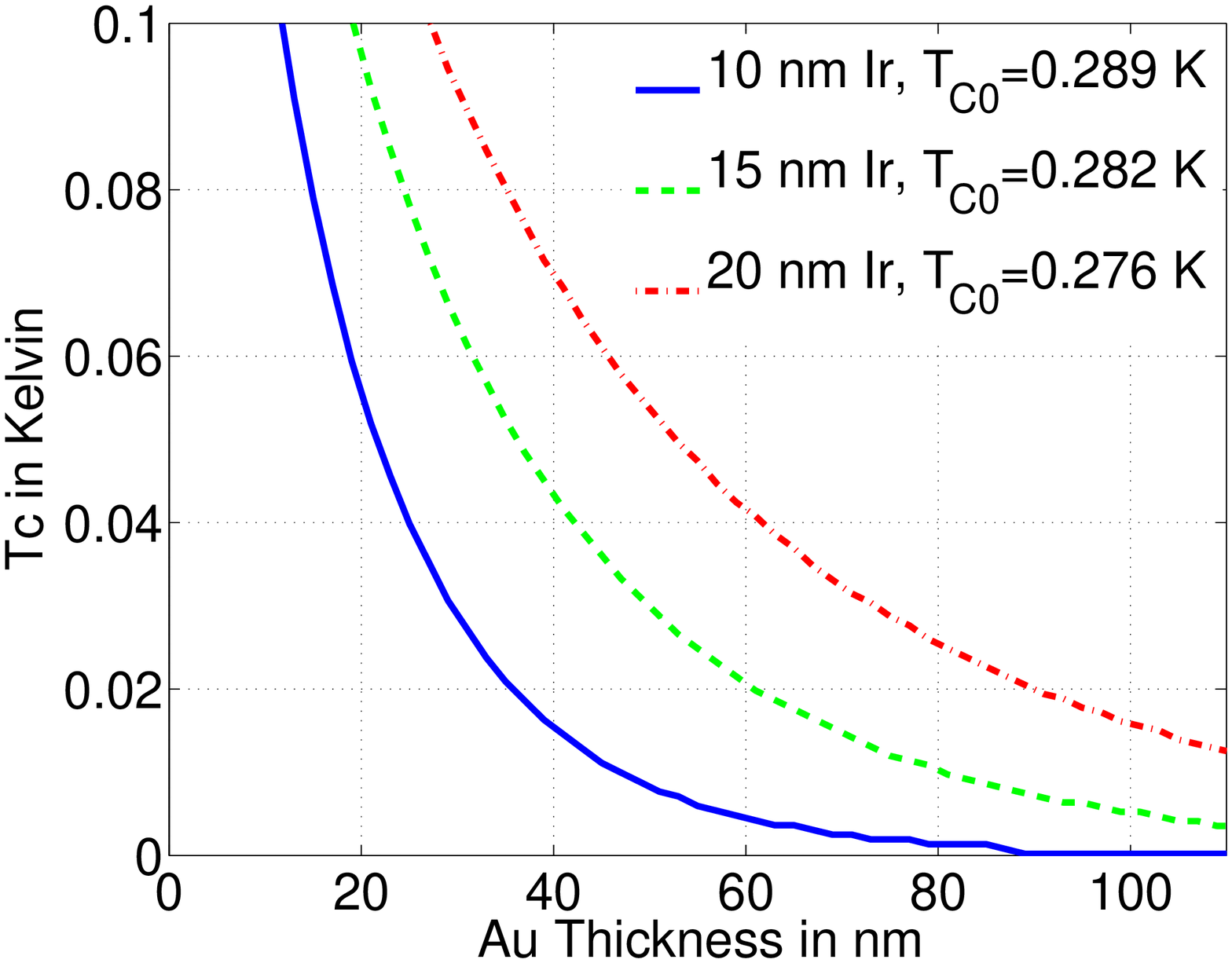}
	\caption{$T_c$ modeling of thin Ir/Pt and Ir/Au bilayers~\cite{Wang:2022}. {\it Left}: $T_c$ vs Pt film thickness for thin Ir/Pt bilayers. {\it Right}: $T_c$ vs Au film thickness for thin Ir/Au bilayers. $T_{C0}$ of a thin Ir film is estimated with McMillan model~\cite{McMillan:68, Bogorin:08} of $T_{C0}=T_0+Ae^{-t/t_0}$, where $t$ is an Ir film thickness, $T_0$=0.121 K, $A$=0.182 K and $t_0$=124.5 nm are extracted with measured data of $T_{C0}$=0.276, 0.253 and 0.203 K with Ir films of 20, 40 and 100 nm respectively.}
	\label{fig:tc}
\end{figure}

\begin{figure}[!htb]
	\centering
	\includegraphics[height=2.2in, keepaspectratio]{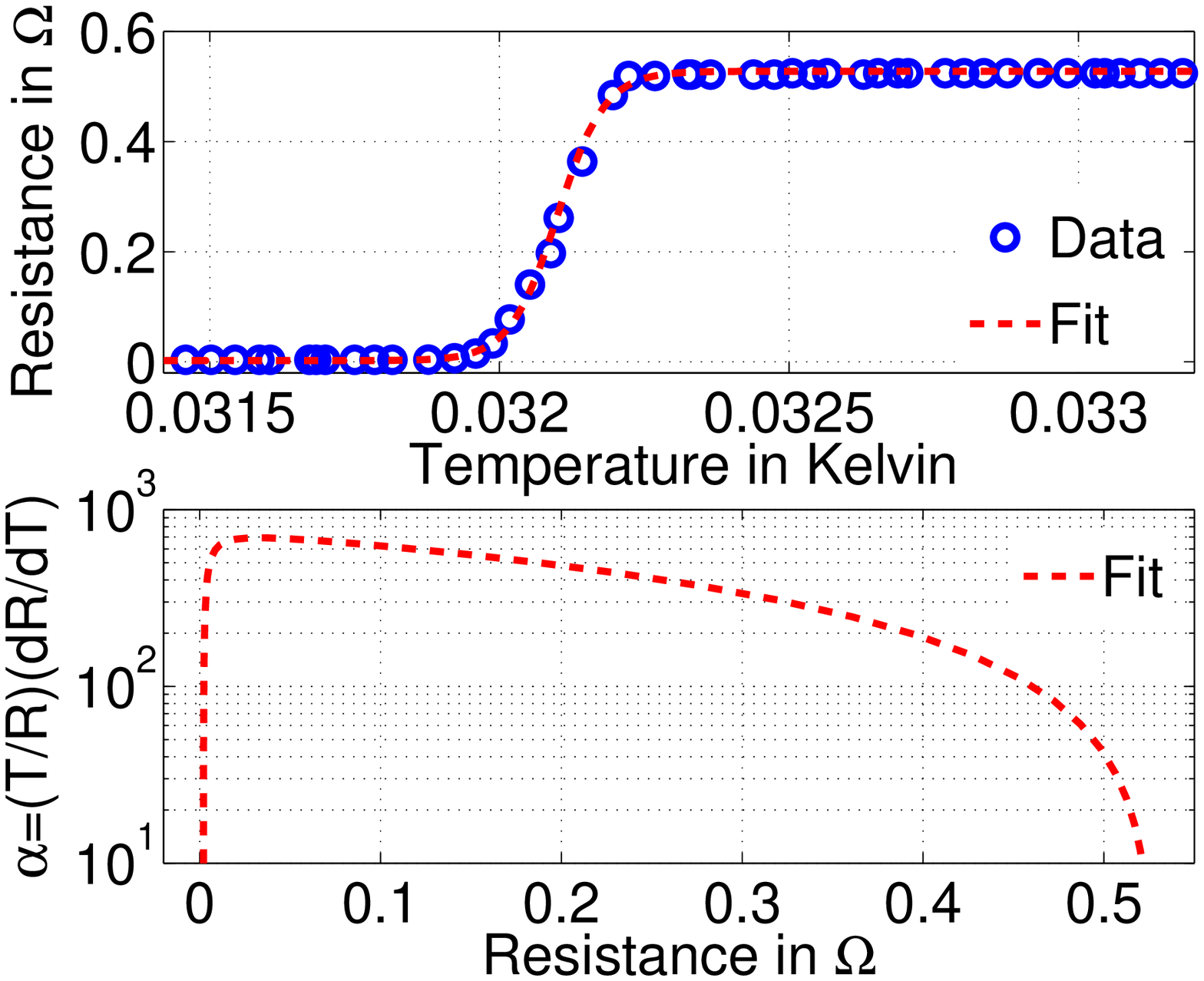}
	\includegraphics[height=2.2in, keepaspectratio]{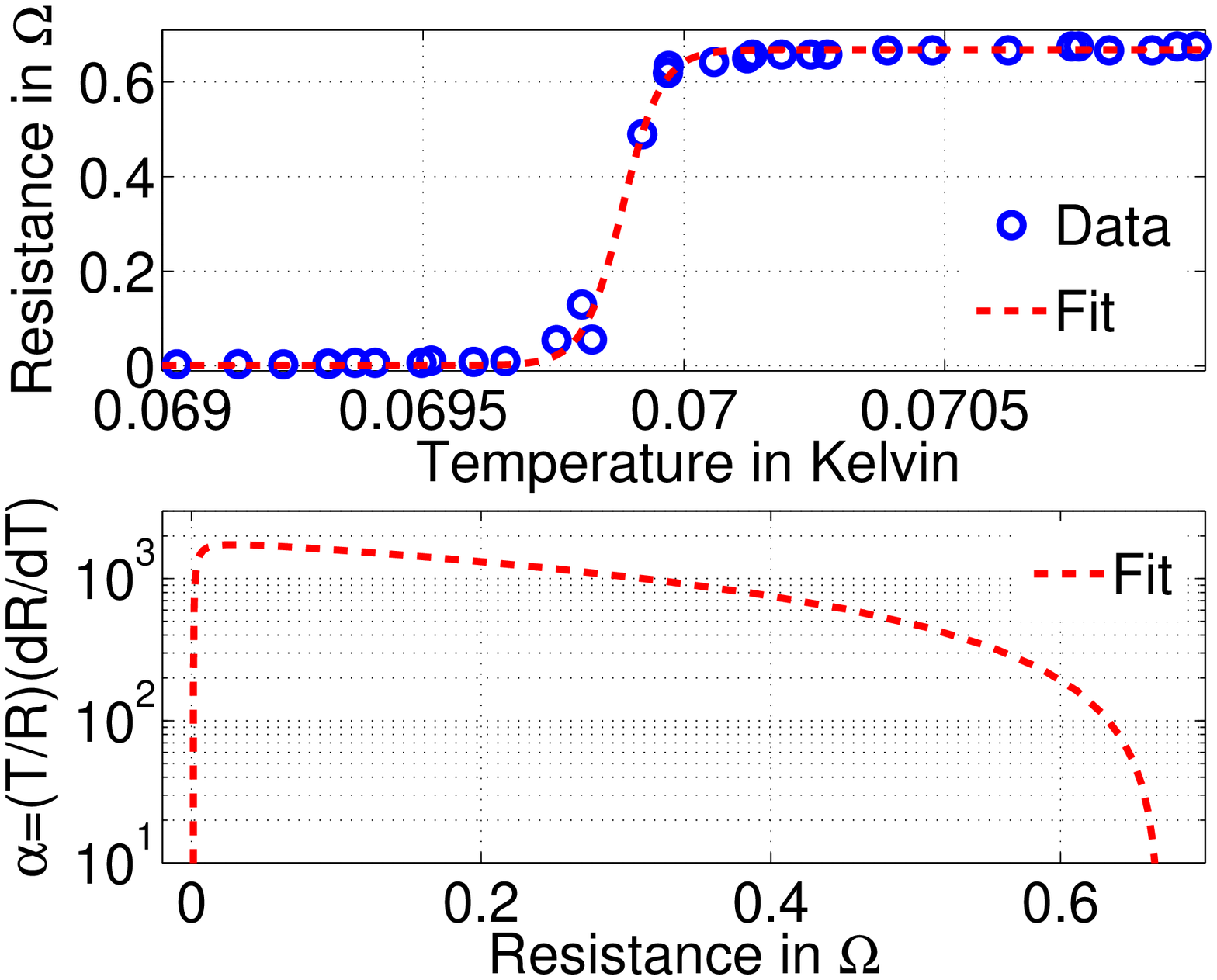}
	\caption{Superconducting-to-resistive transition profiles of Ir/Pt bilayer TES sensors~\cite{Wang:2022}. {\it Left}: R vs T and estimated $\alpha$ vs R of a 500 $\mu$m $\times$ 500 $\mu$m, 100 nm Ir / 80 nm Pt bilayer TES on silicon. $T_c$=32.21 mK. The transition width from 10\% to 90\% $R_n$ is 0.18 mK. {\it Right}: R vs T and estimated $\alpha$ vs R of a 100 $\mu$m $\times$ 100 $\mu$m, 100 nm Ir / 40 nm Pt bilayer TES on silicon. $T_c$=69.87 mK. The transition width from 10\% to 90\% $R_n$ is 0.14 mK. Both of them were made with sputtering at room temperature and lift-off patterning. However, 3N Ir target was used for the TES with data on the left, 4N Ir target was used for the other TES. The R-T curves were measured with Star Cryoelectronics SQUID at a TES current $<$ 10 nA.}
	\label{fig:2}
\end{figure}

The measured Ir-based bilayers and trilayers~\cite{Hennings:20} are thick. 
This is because all of them used 100 nm thick Ir films.
To reduce the TES heat capacity, it makes sense to fabricate thin Ir-based bilayers for low-$T_c$ TES. 
With the proximity models and materials data in reference~\cite{Wang:17}, as well as the extracted electron interface transparencies (0.094 between Ir and Pt and 0.105 between Ir and Au) and electron spin relaxation time in Pt (0.076 ns) in reference~\cite{Hennings:20}, we show recipes for thin Ir/Pt and Ir/Au bilayers with tunable low-$T_c$ in Fig.~\ref{fig:tc}. 
For a thin Ir/Pt bilayer with a $T_c$ of 20 mK, Ir film can be around 30 nm. 
The Pt film thickness can be found around 14 nm by device fabrication and measurements. 
Similarly, for a thin Ir/Au bilayer with a $T_c$ of 20 mK, Ir film should be selected between 10 and 15 nm. 
The Au film thickness can be found between 35 and 65 nm through experiments.
Note that the conductivity of Au film is about 2.5 times that of Ir film and 5 times that of Pt film. 
So, the impedance of a thin Ir/Au bilayer TES can be much smaller than that of a similar Ir/Pt bilayer TES. 

Fig.~\ref{fig:2} shows the record-sharp transition profiles of two patterned Ir/Pt bilayer TES devices fabricated and tested at Argonne.
The transition profiles were determined by a least squares fit of the measured
R-T data to an empirical equation of
\begin{equation} 
R(T)=\frac{R_n}{1+e^{(AT+B)}}+D, 
\label{eq:profile} 
\end{equation} 
where $T$ is the mixing chamber plate temperature, $R$ is the measured resistance as a function of $T$, $R_n$ is the normal resistance above $T_c$, and $D$ is the parasitic resistance. 
The critical temperature is evaluated at $R$ = 50\% of $R_n$, $T_c=-B/A$.
The $\alpha$ in Fig.~\ref{fig:2} is calculated with the fit function in equation~\ref{eq:profile}. 
An Ir/Pt bilayer TES has a sharp transition profile, therefore, a large $\alpha$ for an excellent energy resolution.

\begin{figure}[!htb]
    \begin{center}
    \includegraphics[height=2.0in, keepaspectratio]{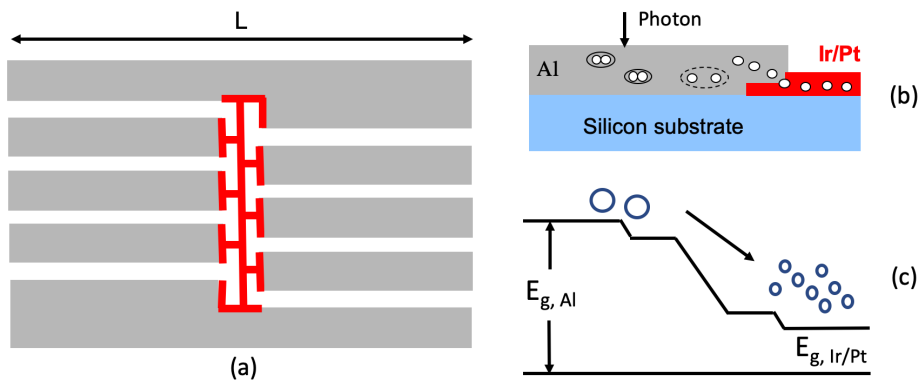} 
    \caption{Principle of quasiparticle generation and propagation~\cite{Wang:2022}. (a) Top view of a QET made of Al photon collection fins and an Ir/Pt bilayer TES. (b) Illustration of quasiparticle generation and propagation after a photon hits a Al fin. (c) Quasiparticle diffusion driven by chemical potential difference between Al and Ir/Pt and quasiparticle number amplification because of energy gap difference betweemn Al and Ir/Pt bilayer.}
    \end{center}
    \label{fig:quasiparticle}
\end{figure}

A more advanced technology to further reduce the TES heat capacity for a higher resolution is to use Quasiparticle-trap-assisted Electro-thermal-feedback Transition-edge-sensor (QET)~\cite{Cabrera:00, Saab:02}. Fig.~\ref{fig:quasiparticle}(a) is the top view of a QET structure, which consists of Al fins and an Ir/Pt TES. 
Its operational principle is the following.  
Photons (or phonons) break cooper pairs in superconducting aluminum fins.  
The unpaired electrons, also called quasiparticles, diffuse into Ir/Pt TES driven by chemical potential difference, and produce more quasiparticles. 
See Fig.\ref{fig:quasiparticle}(b) and (c).  
These injected quasiparticles change Ir/Pt TES resistance. 
For a QET, the volume of the TES is effectively reduced for sensitivity enhancement. 
Therefore, this is an advanced TES technology with an exceptional sensitivity. 
One key parameter of a QET is the quasiparticle diffusion length $l$ in Al. 
The diffusion length $l$ is 135 $\mu$m for a sputtered 300 nm Al film~\cite{Yen:14} and 180 $\mu$m for a sputtered 350 nm Al film~\cite{Pyle:06}.
It is up to 1.5 mm for a evaporated 1 $\mu$m Al film~\cite{Loidl:01}. 
The diffusion length of quasiparticle in Al film strongly depends on the film thickness.
The length $L$ in Fig.~\ref{fig:quasiparticle}(a) should be less than the diffusion length $l$ for a high-resolution TES detector with QET structures.
Future research of QET may include Al film fabrication quality control, the effect of Al film thickness on quasiparticle diffusion length, and quasiparticle interface transparency between Al film and Ir/Pt bilayer. 

\section{Conclusion}
\label{Conclusion}
Hydrogen-rich crystals, which include water ice and selected hydrocarbons, can be ideal detection targets for light DM searches. 
The kinematics of scattering of a sub-GeV DM particle off the hydrogen nucleus results in the maximum kinetic energy deposition possible.
The readout signals are mid-infrared photons and athermal phonons, both of which take low excitation energy. 
The enabling technology to detect single mid-infrared photons or a small flux of athermal phonons is low-$T_c$ TES detectors which are ultra-sensitive. 
Therefore, utilization of hydrogen-rich crystals and low-$T_c$ TES detectors allows to measure the unexplored parameter space of light DM particles. 
In the high energy physics community, there are three major pieces of work to do for using hydrogen-rich crystals to detect light DM.
First, comparison of various hydrogen-rich target crystals in terms of science reach is on demand.
The spin-dependent (-independent) nucleon scattering cross sections for sub-GeV DM should be interpreted in terms of the effective field theory~\cite{fitzpatrick:13, Gresham:14} in DM direct detection.
Second, the detection threshold of low-$T_c$ detectors should be as low as possible.
The R\&D work includes TES materials with a lower $T_c$ and a sharper transition profile, detector layout optimization, heat mass control, and use of a QET structure for large area coverage. 
Third, technologies directly related to a DM search experiment with hydrogen-rich crystals should be investigated.
The relevant studies include the mean free paths of the infrared photons and the athermal phonons in the target crystals, as well as optimal detection readout schemes such as measuring infrared photons with thin film crystals, or athermal phonons with bulk crystals, or both infrared photons and athermal phonons with bulk crystals.

\section*{Acknowledgement}
The work at the Argonne National Laboratory, including the use of facility at the Center for Nanoscale Materials, was supported in part by the Office of Science and in part by the Office of Basic Energy Sciences of the U.S. Department of Energy under Contract DE-AC02-06CH11357. 

\bibliographystyle{unsrt}  
\bibliography{main}  

\end{document}